\documentclass[12pt,a4paper,final]{iopart}
\usepackage{iopams}  
\usepackage{graphicx}
\usepackage[breaklinks=true,colorlinks=true,linkcolor=blue,urlcolor=blue,citecolor=blue]{hyperref}
\usepackage{cases}

\begin{document}

\title[Appropriate probe condition for absorption imaging of ultracold $^6$Li atoms]{Appropriate probe condition for absorption imaging of ultracold $^6$Li atoms}

\author{Munekazu Horikoshi$^{1,2}$, Aki Ito$^3$, Takuya Ikemachi$^4$, Yukihito Aratake$^4$, Makoto Kuwata-Gonokami$^{1,2,4}$, and Masato Koashi$^{1,2,3}$}
\address{$^1$Institute for Photon Science and Technology, Graduate School of Science, The University of Tokyo, 7-3-1 Hongo, Bunkyo-ku, Tokyo 113-8656, Japan.}
\address{$^2$Photon Science Center, Graduate School of Engineering, The University of Tokyo, 2-11-16 Yayoi, Bunkyo-ku, Tokyo 113-8656, Japan.}
\address{$^3$Department of Physics, Graduate School of Science, The University of Tokyo, 7-3-1 Hongo, Bunkyo-ku, Tokyo 113-0033, Japan.}
\address{$^4$Department of Applied Physics, Graduate School of Engineering, The University of Tokyo, 7-3-1 Hongo, Bunkyo-ku, Tokyo 113-8656, Japan.}

\begin{abstract}
One of the readily accessible observables in trapped cold-atom experiments is the column density, which is determined from optical depth (OD) obtained from absorption imaging and the absorption cross-section ($\sigma_{\rm abs}$).
Here we report on simple and accurate determination of OD for dense gases of light atoms such as lithium-6.
We investigate theoretically and experimentally an appropriate condition for the probe intensity and duration to achieve good signal-to-noise ratio by considering the influences of photon recoils and photon shot noises.
As a result, we have succeeded in measuring OD which reached 2.5 with a signal-to-noise ratio of 10 under spatial resolution of 1.7~$\mu$m.
\end{abstract}

\section{Introduction}

Studies of cold atoms contribute to many different areas of physics. One example is the investigation of strongly interacting Fermi gases in the Bardeen-Cooper-Schrieffer to Bose-Einstein condensation (BCS-BEC) crossover using Feshbach resonances \cite{ref1,ref2,ref2.5}, where fermions interact with an $s$-wave scattering length at ultracold temperatures. Such a Fermi system is an ideal simulator for dilute neutron matter, which exists in neutron stars, neutron-rich nuclei, and heavy ion collisions. When the equation of state (EOS) of such interacting fermions is determined using cold atoms, it will provide information regarding many-body physics in dilute nuclear matter.

One of the important parameters to characterize the low temperature properties of Fermi systems is $T/T_F$, which is temperature normalized according to the Fermi temperature.
It is advantageous to use dense fermions for the investigation of many-body physics at the zero-temperature limit because $T/T_F\propto Tn^{-2/3}$, where $n$ is the number density.
One standard technique used to observe atomic density is absorption imaging.
This technique has been successfully used to observe variety of cold atoms, for example, single atom \cite{ref4.5}, mesoscopic atomic clouds \cite{ref4,ref15.5}, dense atomic clouds \cite{ref5}, Bose-Einstein condensates \cite{ref3}, degenerate Fermi gases \cite{ref2.5, ref4.6}, and direct absorption imaging of ultracold polar molecules \cite{ref4.7}. 
However, there have been some technical issues when we would like to observe dense gases of light atoms as discussed below.

When a probe laser with intensity $I$ is applied to atoms along the $y$ direction, the probe intensity decreases according to the Beer-Lambert law with correction for saturation \cite{ref4,ref5,ref13.7}:
\begin{equation}
\frac{dI(x,y,z)}{dy}=-n_{\rm 3D}(x,y,z)\frac{\sigma_{\rm abs}}{1+\frac{I(x,y,z)}{I_{\rm sat}}}I(x,y,z),
\label{eq1}
\end{equation}
where $\sigma_{\rm abs}$ and $I_{\rm sat}$ are respectively the absorption cross-section and saturation intensity.
In this equation, it is assumed that the atoms keep the resonant condition and constant density during measurement.
Here we define the incident and output probe intensities as $I_{\rm in}(x, z)=I(x,y=-\infty,z)$ and $I_{\rm out}(x, z)=I(x,y=+\infty,z)$, the saturation parameter as $s(x, z)=I_{\rm in}(x, z)/I_{\rm sat}$, the transmittance as $T_{\rm abs} (x,z)=I_{\rm out} (x,z)/I_{\rm in} (x,z)$, and the column density as ${\bar n}(x,z)=\int_{-\infty}^{+\infty}n_{\rm 3D}(x,y,z)dy$.
The column density is derived from Eq.~(\ref{eq1}) as
\begin{equation}
{\bar n}(x, z)=\frac{1}{\sigma_{\rm abs}}\left\{-{\rm log}(T_{\rm abs}(x, z))+s(x, z)\cdot(1-T_{\rm abs}(x, z))\right\}.
\label{eq2}
\end{equation}
Therefore, a set of $T_{\rm abs}(x, z)$, $s(x, z)$, and $\sigma_{\rm abs}$ is required to obtain a spatial distribution of the column density.
In this paper, optical depth (OD) is defined as $OD(x, z)=\sigma_{\rm abs}\cdot {\bar n}(x, z)$.

It is straightforward to evaluate $T_{\rm abs}$; 
however, the Beer-Lambert law (\ref{eq1}) can be violated under inappropriate probe conditions.
During irradiation with a probe pulse, atoms repeat cycles of absorption and spontaneous emission, whereby atoms accumulate velocity changes caused by photon recoils.
This induces detuning of the probe laser due to the Doppler shift and blurring induced by random walk \cite{ref4.6}.
Furthermore, $I_{\rm sat}$ and $\sigma_{\rm abs}$ often have values different from their theoretical values for various reasons, such as the line width and polarization of the probe laser, and imperfect closed imaging transition \cite{ref5}.

A weak probe pulse ($s\ll 1$) is typically applied to atoms with a short pulse duration to suppress the influence of the Doppler shift and random walk.
In this case, OD can be approximated by the first term of Eq.~(\ref{eq2}), and the value of $I_{\rm sat}$ does not have to be evaluated.
However, this condition is not suitable for imaging a dense cloud with a large magnification, because a small number of photons per charge-coupled device (CCD) pixel gives a poor signal-to-noise ratio (SNR).

Imaging of a dense cloud of $^{87}$Rb with an intense probe pulse ($s>1$) has been previously demonstrated~\cite{ref5}.
However, the issues of Doppler shift and random walk due to frequent photon recoils during the intense probe pulse irradiation were not discussed.
While the influence of photon recoils is almost negligible for heavy atomic species such as $^{87}$Rb, it becomes significant for light $^6$Li due to a larger recoil velocity.
$I_{\rm sat}$ and $\sigma_{\rm abs}$ were also calibrated using a common calibration coefficient~\cite{ref5}.
As discussed in Ref.~\cite{ref5}, the accuracy of the calibration coefficient depends on the accuracy of determining the probe laser intensity applied to the atoms.
Since there are several parameters used to evaluate probe intensity at the atomic position, it is sometimes difficult to evaluate them with high accuracy.

In this paper, we report on a simple model for the appropriate probe condition to measure  column density $\bar{n}$ of a dense gas of ultracold $^6$Li atoms accurately and show the experimental examination.
In Sec.~\ref{absorption}, we describe the general theory of absorption imaging in terms of counts of a CCD camera.
In Sec.~\ref{probecondition}, we propose a model for the appropriate probe condition by considering the influences of photon recoils and the photon shot noises.
In Sec.~\ref{saturationparameter}, we examine the model experimentally.
The conclusions of this work are given in Sec.~\ref{summary}.
We also provide several experimental techniques in the appendix.
In \ref{Imaging}, we explain our imaging system which enables to take images of two spin states of $^6$Li simultaneously in the magnetic field for the Feshbach resonance with a spatial resolution of 1.7~$\mu$m.
In \ref{crosssection}, we introduce principle to calibrate $\sigma_{\rm abs}$ by using local pressure of an ideal Fermi gas and the equation of state.

\section{\label{absorption}Absorption imaging}

A two-dimensional signal that is proportional to the probe intensity $I(x, z)$ at the object plane on which the atoms are trapped can be acquired using a CCD camera.
When the probe intensity is $I$ and the probe duration is $t$, the CCD count at the corresponding pixel is given by
\begin{equation}
C(I, t)=N_{{\rm e}^-}(I, t) \cdot G_{\rm CCD},
\label{CCDcount}
\end{equation}
where $N_{{\rm e}^-}(I, t)=N_{\rm p}(I, t)Q_{\rm e}$ is the number of photo-electrons (PE), $N_{\rm p}(I, t)=\frac{I L_{\rm pix}^2 t}{E_{\rm p}}T_{\rm p}$ is the number of photons hitting the pixel during $t$, $E_{\rm p}=h c/\lambda$ is the photon energy, and $T_{\rm p}$ is the transmittance of the probe laser from inside the glass cell to the CCD camera. $Q_{\rm e}$ is the quantum efficiency of the CCD camera, and $G_{\rm CCD}=Gain/Sensitivity$ is a parameter determined by the CCD gain for PEs and the CCD sensitivity, which has units of PEs/count.
Let $(i, j)$ be the address of the two-dimensional arrays of pixels at the imaging plane, and $(x_i, z_j)$ be the corresponding position at the object plane.
Here the CCD counts for input and output are defined as $C_{\rm in}(i, j)=C\left(I_{\rm in}(x_i, z_j), t\right)$ and $C_{\rm out}(i, j)=C\left(I_{\rm out}(x_i, z_j), t\right)$, respectively.
The count that corresponds to the saturation intensity is also defined as $C_{\rm sat}=C(I_{\rm sat}, t)$.

Under realistic experimental conditions, the CCD count includes background noise, such as from stray light and CCD noise.
To cope with this noise, three images are acquired of the probe laser with atoms ($C_{\rm abs}$), the probe laser without atoms ($C_{\rm probe}$), and the background without the probe laser ($C_{\rm back}$), and the background is removed accordingly:
\begin{eqnarray}
C_{\rm in}(i, j)=C_{\rm probe}(i, j)-C_{\rm back}(i, j), \label{eq14} \\
C_{\rm out}(i, j)=C_{\rm abs}(i, j)-C_{\rm back}(i, j). \label{eq15}
\end{eqnarray}
$C_{\rm out}(i, j)$ sometimes takes unphysical negative values due to low SNR.
We omitted such pixels in the subsequent data analysis.

Since $C(I, t)\propto I$, the transmittance and the saturation parameter are given by $T_{\rm abs}(x_i, z_j)=\frac{C_{\rm out}(i, j)}{C_{\rm in}(i, j)}$ and $s(x_i, z_j)=\frac{C_{\rm in}(i, j)}{C_{\rm sat}}$, respectively.
We divide $C_{\rm sat}(t)$ into the time-independent part $\chi_{\rm sat}$ and $t$:
\begin{equation}
C_{\rm sat}(t)=\chi_{\rm sat}\cdot t.
\label{eq5}
\end{equation}
In this paper, we call $\chi_{\rm sat}$ saturation constant.
The OD given by Eq.~(\ref{eq2}) can then be expressed using these CCD counts as
\begin{equation}
OD(x_i, z_j)=-{\rm log}\left( \frac{C_{\rm out}(i, j)}{C_{\rm in}(i, j)} \right)+\frac{C_{\rm in}(i, j)-C_{\rm out}(i, j)}{\chi_{\rm sat}\cdot t}.
\label{eq6}
\end{equation}

We define two values as follows:
\begin{eqnarray}
C_1(i, j)=-{\rm log}\left( \frac{C_{\rm out}(i, j)}{C_{\rm in}(i, j)} \right), \label{eq19} \\
C_2(i, j)=\frac{C_{\rm in}(i, j)-C_{\rm out}(i, j)}{t}. \label{eq20}
\end{eqnarray}
Then Eq.~(\ref{eq6}) can be transformed to
\begin{equation}
C_1(i, j)=OD(x_i, z_j)-\frac{C_2(i, j)}{\chi_{\rm sat}}.
\label{eq21}
\end{equation}
If $C_1$ and $C_2$ are measured at various probe intensities, then they must obey the linear law.
Therefore, the value of $\chi_{\rm sat}$ can be easily determined from the gradient of $C_1$ relative to $C_2$ without knowing each parameter that composes $\chi_{\rm sat}$.
This is the principle of the proposed method to determine $\chi_{\rm sat}$.

\section{\label{probecondition}Appropriate probe conditions}

All of the formulas discussed above are based on the Beer-Lambert law given by Eq.~(\ref{eq1}), which is established under the condition that atoms maintain a resonant condition and constant density during the probe pulse duration.
In addition, the measured SNR of the OD should be sufficiently high.
We would like to find the appropriate probe condition in order to use Eq.~(\ref{eq1}) without using additional experimental techniques.
Thus we consider a model including the following three conditions for the appropriate probe laser pulse, and we examine the validity of the model experimentally.

\subsection{Doppler condition}

When the probe laser has detuning, which is defined as $\delta=\frac{\omega_L-\omega_0}{\Gamma/2}$, Eq.~(\ref{eq1}) is modified to \cite{ref4}
\begin{equation}
\frac{dI(x,y,z)}{dy}=-n_{\rm 3D}(x,y,z)\frac{\sigma_{\rm abs}}{1+\frac{I(x,y,z)}{I_{\rm sat}}+\delta^2}I(x,y,z),
\label{eqmod1}
\end{equation}
where $\omega_L$, $\omega_0$, and $\Gamma$ are the laser angular frequency, the resonant frequency, and the natural linewidth of the atomic transition.
The detuning caused by the Doppler shift is $\delta_D=\frac{\omega_D}{\Gamma/2}=\frac{v_{\rm rec} k}{\Gamma/2}$ per photon recoil, where $v_{\rm rec}=\hbar k/m$ is the recoil velocity of atoms with the mass $m$ for the probe laser with the wave number $k$.
Then the maximum Doppler shift is given by $\delta_D(t, s)=N_{\rm sc}(t, s)\frac{\omega_D}{\Gamma/2}$.
$N_{\rm sc} (t,s)=R_{\rm sc} (s)t$ is the mean number of photon scattering events, which
is estimated from the scattering rate, which is given by $R_{\rm sc} (s)=\frac{\Gamma}{2}\frac{s}{1+s}$.

Since the detuning changes in time during applying the probe laser pulse, it is not simple to find the probe condition where Eq.~(\ref{eq1}) is approximately satisfied.
Here, we simply set the condition of $1+\frac{I}{I_{\rm{sat}}} > \delta_D^2(t,s)$ as a criterion for suppressing influence of $\delta$ in Eq.~(\ref{eqmod1}), that is,
\begin{equation}
\sqrt{1+s} > N_{\rm sc}(t, s)\frac{\omega_D}{\Gamma/2}.
\label{eq22}
\end{equation}
We call this inequality Doppler condition. 
This explicit form is given by
\begin{equation}
t<\frac{m\lambda^2}{2\pi h}\frac{(1+s)^{3/2}}{s}.
\label{eqDoppler}
\end{equation}
Equation~(\ref{eqDoppler}) shows that the Doppler condition is dependent on $m$ and is not dependent on $\Gamma$.
Therefore, light atoms are more susceptible than heavy atoms to the influence of the Doppler effect caused by photon recoils.
The Doppler conditions for $^6$Li ($\lambda=671$~nm) and $^{87}$Rb ($\lambda=780$~nm) are shown by the dotted red curves in Fig.~\ref{fig3}.
The regions below the red dotted curves satisfy the Doppler condition.

\subsection{Random walk condition}

The second condition is the random walk condition to ensure atoms stay within an area $S=L_{\rm pix}^2$ during the probe pulse.
This condition is given by 
\begin{equation}
\delta r_{\rm rec}(t,s)<L_{\rm pix}.
\label{eq23}
\end{equation}
The left-hand side term is the displacement after random walk on the object plane, which is estimated to be $\delta r_{\rm rec} (t,s)=\int_0^t\sqrt{\langle v_{\rm N}^2 \rangle (t',s)}dt'$ using a speed given by $\langle v_{\rm N}^2 \rangle(t,s)=\frac{2}{3} v_{\rm rec}^2 N_{\rm sc} (t,s)$ \cite{ref3,ref13,ref13.5}.
Thus, the explicit form of Eq.~(\ref{eq23}) is given by
\begin{equation}
t<\frac{3}{2^{2/3}}\left( \frac{m\lambda}{h}L_{\rm pix}\sqrt{\frac{1}{\Gamma}\frac{1+s}{s}} \right)^{2/3}.
\label{eqRandom}
\end{equation}
Equation~(\ref{eqRandom}) shows that the random walk condition is dependent on $m^{2/3}$.
Therefore, light atoms are more influenced by random walk caused by photon recoils than heavy atoms.

For our experimental condition, the size of the visual field per pixel is $L_{\rm pix}=1.7$~$\mu$m (see \ref{Imaging}).
The random walk conditions under our imaging condition are shown for $^6$Li ($\lambda=671$~nm, $\Gamma=2\pi \times 5.87$~MHz) and $^{87}$Rb ($\lambda=780$~nm, $\Gamma=2\pi \times 6.07$~MHz) by the dashed blue curves in Fig.~\ref{fig3}.
The regions below the blue dashed curves satisfy the random walk condition.

\subsection{SNR condition}

The third condition is the SNR condition for the signal strength.
The SNR is required to be larger than unity:
\begin{equation}
SNR(s,t,OD)>1.
\label{eq24}
\end{equation}
The SNR of the OD is given by the mean value and variance of OD as
\begin{equation}
SNR(s,t,OD)=\frac{OD}{\sqrt{\sigma_{OD}^2(s,t,OD)}},
\label{eq13}
\end{equation}
where the variance of $OD$ is given by
\begin{eqnarray}
\sigma_{\rm OD}^2(s,t,OD)&=&\left( \frac{\partial OD}{\partial C_{\rm in}} \right)^2\sigma_{C_{\rm in}}^2(s,t)+\left( \frac{\partial OD}{\partial C_{\rm out}} \right)^2\sigma_{C_{\rm out}}^2(s,t,OD)\nonumber \\
&=&(1+s)^2\left( \frac{\sigma_{C_{\rm in}}}{\langle C_{\rm in}\rangle} \right)^2+\left(1+sT_{\rm abs}\right)^2\left( \frac{\sigma_{C_{\rm out}}}{\langle C_{\rm out}\rangle} \right)^2.
\label{eq12}
\end{eqnarray}
The transmittance $T_{\rm abs}$ is given as a function of $s$ and $OD$ by solving Eq.~(\ref{eq2}):
\begin{equation}
T_{\rm abs}(s,OD)=s^{-1}W[s~{\rm exp}(s-OD)],
\label{eq9}
\end{equation}
where $W$ is Lambert's $W$ function \cite{ref4, ref13.7}.

Let us define the number of PEs for $I=I_{\rm sat}$ as
\begin{equation}
N_{\rm sat}(t)=N_{e^-}(I_{\rm sat},t)=\frac{I_{\rm sat}L_{\rm pix}^2t}{hc/\lambda}T_{\rm p}Q_{\rm e},
\label{Nsat}
\end{equation}
from which the mean values of $C_{\rm in}$ and $C_{\rm out}$ can be described as
\begin{eqnarray}
\langle C_{\rm in}(s,t)\rangle=sN_{\rm sat}(t) \cdot G_{\rm CCD}, \label{eq7} \\
\langle C_{\rm out}(s,t,OD)\rangle=sN_{\rm sat}(t)T_{\rm abs}(s,OD)\cdot G_{\rm CCD}. \label{eq8}
\end{eqnarray}
The counts $\langle C_{\rm in}\rangle$ and $\langle C_{\rm out}\rangle$ are experimentally determined through the relations in Eqs.~(\ref{eq14}) and (\ref{eq15}).
In the present experiment, the background is not dependent on the probe duration $t$, because the pulse duration $t$ is changed under a fixed CCD exposure time.
Denoting this constant value as $N_{\rm back} G_{\rm CCD}$, we have $\langle C_{\rm back} \rangle=N_{\rm back}G_{\rm CCD}$, $\langle C_{\rm probe}\rangle=\langle C_{\rm in}\rangle+N_{\rm back}G_{\rm CCD}$, and $\langle C_{\rm abs}\rangle=\langle C_{\rm out}\rangle+N_{\rm back}G_{\rm CCD}$.
When only the photon shot noise and quantum efficiency $Q_{\rm e}$ are considered, and the atomic shot noise and the noise factor caused by the multiplication processes of PEs in the CCD camera are ignored \cite{ref12}, then the variances of $C_{\rm in}=C_{\rm probe}-C_{\rm back}$ and $C_{\rm out}=C_{\rm abs}-C_{\rm back}$ can be described as
\begin{eqnarray}
\sigma_{C_{\rm in}}^2(s,t)=\left(sN_{\rm sat}(t)+2N_{\rm back}\right) \cdot G_{\rm CCD}^2, \label{eq10} \\
\sigma_{C_{\rm out}}^2(s,t,OD)=\left(sN_{\rm sat}(t)T_{\rm abs}(s, OD)+2N_{\rm back}\right) \cdot G_{\rm CCD}^2. \label{eq11}
\end{eqnarray}
The SNR of the OD does not appear to depend on the CCD factor $G_{\rm CCD}$, because it is canceled in Eq.~(\ref{eq12}).
The explicit form of Eq.~(\ref{eq13}) is then given by
\begin{eqnarray}
SNR(s, t, OD)=\nonumber\\
\frac{OD\sqrt{sN_{\rm sat}(t)T_{\rm abs}(s, OD)}}{\sqrt{(1+s)^2T_{\rm abs}(s, OD)\left( \frac{2N_{\rm back}}{sN_{\rm sat}(t)}+1 \right)+(1+sT_{\rm abs}(s, OD))^2\left( \frac{2N_{\rm back}}{sN_{\rm sat}(t)T_{\rm abs}(s, OD)}+1 \right)}}.\nonumber\\
\label{SNRexp}
\end{eqnarray}

\begin{figure}[tb!]
 \centering
 \includegraphics{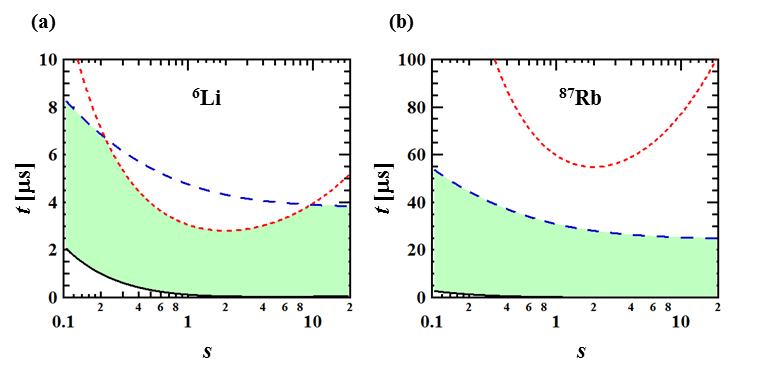}
 \caption{\label{fig3}
Appropriate conditions of the probe laser pulse for (a) $^6$Li and (b) $^{87}$Rb under an effective pixel size of $L_{\rm pix}=1.7$~$\mu$m.
The red dotted curves indicate the boundaries of the Doppler condition calculated using Eq.~(\ref{eqDoppler}).
The blue dashed curves indicate the boundaries of the random walk condition calculated using Eq.~(\ref{eqRandom}).
The black curves show the boundaries of the SNR condition from Eq.~(\ref{SNRexp}) with $OD=2.5$ and $N_{\rm back}=20$.
The shaded green areas represent the appropriate probe conditions.
}
\end{figure}

The SNR condition was calculated for $OD = 2.5$, which was a typical peak OD obtained in the present experiment, using Eq.~(\ref{SNRexp}) with the theoretical values $I_{\rm sat}=I_{\rm sat}^0$ ($I_{\rm sat}^0=2.54$~mW/cm$^2$ for $^6$Li \cite{ref14}, and $I_{\rm sat}^0=1.67$~mW/cm$^2$ for $^{87}$Rb \cite{ref15}). The parameters $T_{\rm p}=0.7$, $Q_{\rm e}=0.75$, $L_{\rm pix}=1.7$~$\mu$m, and $N_{\rm back}=20$ were then experimentally determined.
The SNR condition is shown by the black curves in Fig.~\ref{fig3}.
The regions above the black curves satisfy the SNR condition.
This condition is not dependent on $m$.

\section{\label{saturationparameter}Examination of the model for the appropriate probe condition}

\begin{figure}[]
 \centering
 \includegraphics{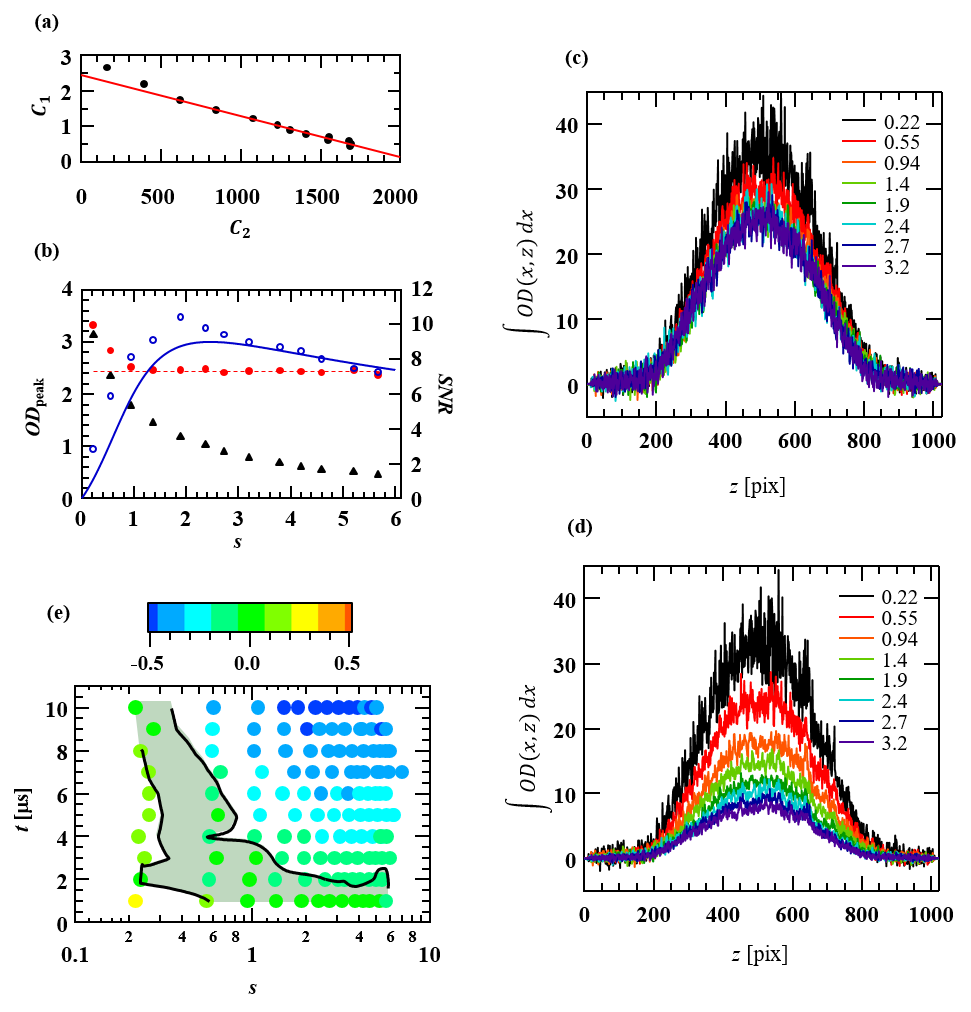}
 \caption{\label{fig4}
Experimental determination of OD and the saturation constant $\chi_{\rm sat}$.
(a) Filled circles show experimental data for $C_1$ as a function of $C_2$ measured at various probe intensities with a fixed pulse duration of 1~$\mu$s.
The solid line is the fitting result with $C_1=\overline{OD}-C_2/\bar{\chi}_{\rm sat}$.
(b) OD and SNR as a function of the saturation parameter around the peak OD.
Filled red circles and the open blue circles show the mean value the OD and SNR, respectively.
The dashed red line shows $\overline{OD}$.
The solid blue curve shows a simulation using Eq.~(\ref{SNRexp}) with $t=1$~ $\mu$s, $OD$=2.5, and $N_{\rm back}=20$.
Black triangles show the OD calculated using only the first term of Eq.~(\ref{eq6}).
(c) Distribution of OD integrated along the $x$ direction measured with the same probe conditions of (a) with the saturation parameter $\bar{\chi}_{\rm sat}$.
(d) Distribution of OD integrated along the $x$ direction measured with the same probe conditions of (a) without considering the second term of Eq.~(\ref{eq6}).
The colors in (c) and (d) indicate the saturation parameters.
(e) Degree of deviation $\Delta OD(t,s)$ from $\overline{OD}$ for various probe conditions.
The filled circles are data points and the color indicates the values.
The solid curves show the borders at $|\Delta OD|=0.1$ and the shaded area satisfies $|\Delta OD| \leq 0.1$.
}
\end{figure}

Here we examine our model given in Eqs.~(\ref{eqDoppler}), (\ref{eqRandom}), and (\ref{eq24}) by checking linearity of  $C_1$ as a function of $C_2$ as expected in Eq.~(\ref{eq21}).
We also compare the SNR of the OD with the theoretical value calculated by Eq.~(\ref{SNRexp}).
Furthermore, we evaluate OD taken by using various probe condition in order to see the boundary of the appropriate probe condition.

Our experiment realizing degenerate Fermi gases are reported in Ref.~\cite{ref9}, and our imaging system is explained in \ref{Imaging}.
In order to collect a pair of $C_1$ and $C_2$ at various probe intensity, we repeatedly acquired absorption images under the same experimental conditions by applying probe pulses in the range of $s \in [0.2, 6]$ with a fixed pulse duration of $t = 1$~$\mu$s.
These probe conditions are within the appropriate probe conditions shown in Fig.~\ref{fig3}(a).
For each individual data set taken at each probe intensity, $C_1(i, j)$ and $C_2(i, j)$ were calculated at each pixel using Eqs.~(\ref{eq14}), (\ref{eq15}), (\ref{eq19}), and (\ref{eq20}), and an average was obtained from 1000 pixels around the peak position of the OD.
The measured relationship between $C_1$ and $C_2$ is plotted in Figure~\ref{fig4}(a).
The data indicate good linearity, as expected from Eq.~(\ref{eq21}), except for two data points obtained using the two weakest probe intensities.
The solid red line shows a linear fitting of the data using Eq.~(\ref{eq21}) with fitting parameters of $OD$ and $\chi_{\rm sat}$.
The two deviated data points were omitted from this fitting procedure.
$OD$ and $\chi_{\rm sat}$ determined by the fitting are denoted as $\overline{OD}$ and $\bar{\chi}_{\rm sat}$, respectively.

In order to evaluate SNR of OD as a function of $s$, we calculated $OD(x_i, z_j)$ and $s(x_i, z_j)$ using the determined $\bar{\chi}_{\rm sat}$ for data taken at each probe intensity.
The mean values $\langle OD \rangle$ and variances $\sigma_{OD}^2$ within the same 1000 pixels were calculated and the SNR was evaluated with respect to $SNR=\frac{\langle OD \rangle}{\sqrt{\sigma_{OD}^2}}$.
Figure~\ref{fig4}(b) shows the mean values (filled red circles) and SNR (open blue circles) of the OD around the peak position as a function of the saturation parameter $s$.
The OD evaluated at various probe intensities have constant values around $\overline{OD}$ (horizontal dashed red line).
The solid blue curve shows a simulation of the SNR calculated using Eq.~(\ref{SNRexp}) with $t=1$~$\mu$ s, $OD$=2.5, and $N_{\rm back}=20$.
The data agree well with the theoretical calculations, which indicates that the SNR can be well explained as photon shot noise of the probe laser beam.
The best SNR was achieved for $s=2$.
At this condition, the number of photon scattering is estimated to be $N_{\rm{SC}}(t=1~\mu s, s=2) \sim 13$.
To reveal the contribution of the second term of Eq.~(\ref{eq6}), the OD calculated using only the first term in Eq.~(\ref{eq6}) is shown by the black triangles.

In principle, the Doppler condition Eq.~(\ref{eqDoppler}) and the Random condition Eq.~(\ref{eqRandom}) do not depend on the atomic density.
In order to confirm that whole OD distribution does not depend on probe intensity, we plot distribution of OD integrated along the $x$ direction in Figure~\ref{fig4}(c).
For comparison, we plot integrated OD calculated using only the first term in Eq.~(\ref{eq6}) in Figure~\ref{fig4}(d).
Figure~\ref{fig4}(c) clearly shows that shape of the integrated OD does not depend on $s$ for $s>0.94$ when OD is calculated with $\chi_{\rm sat}$.
The figure also shows that OD deviates at small $s$ around high atomic density.
From Fig.~\ref{fig4}(a) and Fig.~\ref{fig4}(c), we found that small $s$ is not a suitable probe condition for dense atomic cloud in this experiment.

We have not addressed reason why two points deviate from linearity shown at small $s$ as shown in Fig.~\ref{fig4}(a), and reason why the deviation occurs at high atomic density as shown in Fig.~\ref{fig4}(c).
One of the possible reason is the poor SNR.
Another possibility is imperfection of our model at low probe intensity. 
In any case, we can easily find the appropriate probe condition experimentally by examining the linear region in $C_1(C_2)$ and by evaluating the SNR.

The influence of the Doppler shift and random walk given in Eqs.~(\ref{eqDoppler}) and (\ref{eqRandom}) was examined next.
We again repeatedly acquired absorption images under the same experimental conditions by applying probe pulses with various combinations of intensity and duration were selected within the ranges of $s \in [0.2, 6]$ and $t \in [1, 10]$~$\mu$s.
These probe conditions include those that are outside the appropriate probe conditions shown in Fig.~\ref{fig3}(a).
Data were analyzed with a fixed $\bar{\chi}_{\rm sat}$ in the same procedure as the first experiment.
It was expected that the OD measured under inappropriate conditions would show a deviation from $\overline{OD}$. The deviation was evaluated according to $\Delta OD(t,s)=\frac{OD(t,s)-\overline{OD} }{\overline{OD}}$ and the results are shown in Fig.~\ref{fig4}(e).
The solid circles correspond to the experimental data and the colors show the degree of deviation.
The results clearly show the large influence of the Doppler shift and random walk for probe pulses with long duration and large intensity.
The solid curves show contour plots at $|\Delta OD|=0.1$, which was calculated by using
Igor software package.
The shaded area corresponds to deviation within 10\%, which 
is in agreement with the calculated region shown in Fig.~\ref{fig3}(a).
The boundaries are quantitatively close to the theoretical calculations shown in Fig.~\ref{fig3}(a).
This indicates validity of our model for the appropriate probe condition.

We note that there is another photon recoil effect to distort absorption images.
That is defocussing effect caused by moving atoms with respect to the object plane due to photon recoils \cite{ref4.6,ref15.5}.
The degree of this effect depends on $t$ and $s$.
In our imaging condition as shown in Fig.~\ref{fig4}(b), this effect can be thought to be small, because OD keeps constant values for various $s$.
Also, the appropriate condition determined experimentally as shown in Fig.~\ref{fig4}(e) involves this defocussing effect.

In order to the column density, we have to determine the effective absorption cross-section.
This method has been well established by previous experiments \cite{ref16,ref21}.
In this paper, we introduce several useful technique to determine the cross-section in \ref{crosssection}.

\section{\label{summary}Summary}

In this work, we provided a simple model to estimate the appropriate probe condition (pulse duration and intensity) for absorption imaging.
This model was examined experimentally, and the validity was confirmed.
According to this model, it is straightforward to see that influence of photon recoils is considerable large for light atoms such as $^6$Li in contrast with heavy atoms such as $^{87}$Rb.
By using the optimized probe condition, we succeeded in measuring OD which reached 2.5 with a signal-to-noise ratio of 10 under spatial resolution of 1.7~$\mu$m.
These techniques are powerful tools for the investigation of interacting fermions in the BCS-BEC crossover at the zero-temperature limit.

\section*{Acknowledgement}

The authors are grateful to F\'elix Werner for providing a method of direct transformation from 2D profile to local pressure for an arbitrary trap. This work was supported by Grant-in-Aid for Scientific Research on Innovative Areas (No. 24105006) and a Grant-in-Aid for Young Scientists (A) (No. 23684033).

\appendix

\section{\label{Imaging}Imaging system}

\begin{figure}[tb!]
 \centering
 \includegraphics{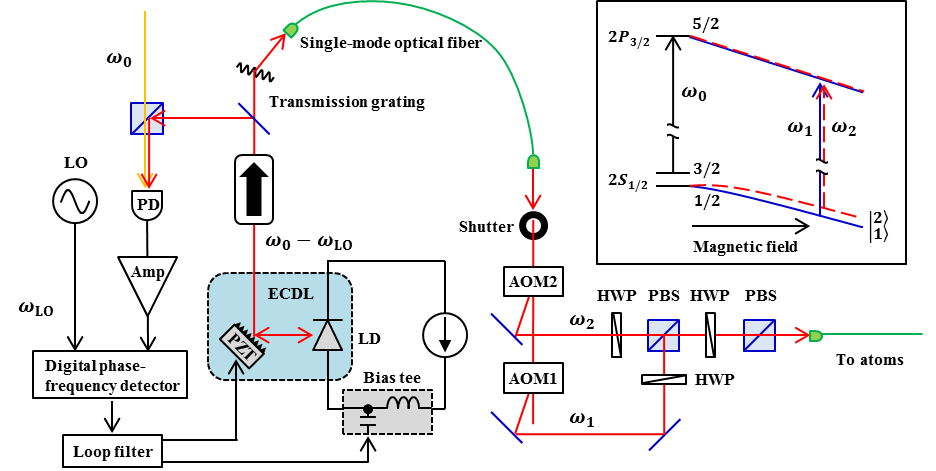}
 \caption{\label{fig1}
Optical system of the probe laser for absorption imaging of $^6$Li atoms. (Inset) Energy diagram for $^6$Li atoms and the transition lines as a function of the magnetic field. The solid blue curves and dashed red curves represent the Zeeman shift of the internal states of $|1\rangle$ and $|2\rangle$, and the corresponding excited states connected by laser light with $\sigma^-$ polarization.
The blue and red arrows show the transition lines used for imaging. 
 (LD: laser diode, PD: photo diode, Amp: rf amplifier, HFP: half-wave plate, PBS: polarizing beam splitter)
}
\end{figure}

Figure~\ref{fig1} shows the optical setup for the probe laser and the inset shows a level diagram of the ground and excited states of $^6$Li atoms in a magnetic field.
$^6$Li atoms have a broad $s$-wave Feshbach resonance at a magnetic field of 832.18~Gauss with a width of 262~Gauss between two internal states: $\left| 1 \right \rangle$~$\equiv$~$\left| m_L= 0,~m_S=-1/2,~m_I=+1\right\rangle$ and $\left| 2 \right \rangle$~$\equiv$~$\left| m_L= 0,~m_S=-1/2,~m_I= 0\right\rangle$ \cite{ref6}.
A mixture of these two internal states shows the BCS-BEC crossover around such a magnetic field at the zero-temperature limit \cite{ref7}.

Absorption imaging is conducted in a high magnetic field (HF-image) to determine the atomic density distribution in such a gas.
Resonant probe lasers are applied to the gas with $\sigma^{-}$ circular polarization. As shown in the inset, $\omega_0$ is defined as the resonant frequency from $F=3/2$ to $F'=5/2$ at zero magnetic field, and $\omega_1$ and $\omega_2$ are respectively defined as the resonant frequencies of $|1\rangle$~$\rightarrow$~$\left| m_L= -1,~m_S=-1/2,~m_I=+1\right\rangle$ and $|2\rangle$~$\rightarrow$~$\left| m_L= -1,~m_S=-1/2,~m_I= 0\right\rangle$ in the magnetic field.
The transition probability is approximately 99.7\% at 832.18~Gauss \cite{ref7.1}, and they are almost closed cycling transitions.

An external cavity diode laser (ECDL) is used as the light source for the HF-image. The output laser is diffracted by a transmission grating and coupled into a single-mode optical fiber to remove amplified spontaneous emission produced by the diode laser. To image $^6$Li in the $|1\rangle$ ($|2\rangle$) state in magnetic fields from 500~Gauss to 1100~Gauss, the frequency $\omega_1$ ($\omega_2$) has to be set with a frequency offset from $-540$ ($-620$)~MHz to $-1.4$ ($-1.5$)~GHz with respect to $\omega_0$ to compensate for the Zeeman shift.
These imaging frequencies are realized using a combination of an optical phase locked loop (OPLL) and two acousto-optic modulators (AOMs). A digital phase-frequency detector (Analog Devices HMC439QS16G) is used in the OPLL system. This device can compare the relative phase and frequency between the two inputs up to 1.3~GHz. The output signal is robust against fluctuation of the input level due the internal comparator and digital processing. These features realize stable OPLL operation with a wide locking range that exceeds 1~GHz. The OPLL is stabilized by feeding the error signal back to the ECDL through a loop filter \cite{ref8}. The two probe laser beams with different frequencies are overlapped with the same polarization and sent to the atoms.
The two AOMs enable first switching of the two probe laser beams.
Then, simultaneous imaging of atoms in the two spin states can be easily accomplished by using one CCD camera operating with the fast shifting mode.

\begin{figure}[tb!]
 \centering
 \includegraphics{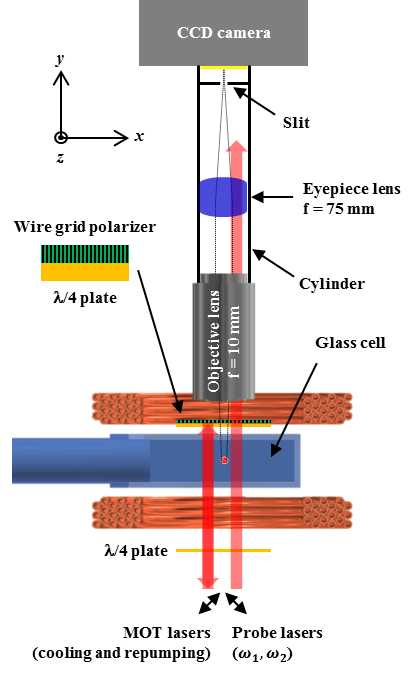}
 \caption{\label{fig2}
Side view of the imaging system. A wire grid polarizer and a $\lambda/4$ plate are mounted on a kinematic mount to align the reflected laser beams for the MOT. A quadrupole magnetic field for the MOT and a bias magnetic field for the Feshbach resonance are produced by the same pair of coils.
}
\end{figure}

Figure~\ref{fig2} shows the imaging setup. $^6$Li atoms are cooled and trapped using a standard magneto-optical trap (MOT) in a vacuum glass cell. They are further cooled by forced evaporative cooling in a hybrid trap consisting of an optical dipole trap (ODT) and a magnetic trap. The ODT is produced by a focused 1070 nm laser beam applied along the $z$ direction. The 1/${\it e}^2$ beam radii of the ODT laser beam are ($w_{0x},w_{0y}$) = (43.5, 46.9)~$\mu$m and the final laser power after evaporative cooling is $P_{\rm ODT}$=45~mW.
The magnetic trap is produced by the magnetic curvature of a bias magnetic field to achieve the Feshbach resonance.
A magnetic curvature is thus produced in the $z$ direction; $\omega_{\rm mag}=2\pi \times 0.24 \sqrt{B}$~Hz for $^6$Li, where $B$ is the produced bias magnetic field in units of Gauss. This magnetic trap works mainly along the $z$ direction because the ODT provides much stronger confinement than the magnetic trap along the $x$ and $y$ directions. Typically, $5\times10^5$ degenerate fermions are prepared in a trap with a depth of 38~$\mu$K and trapping frequencies of ($\omega_x, \omega_y, \omega_z$)=$2\pi \times$($250, 232, 7.08$)~Hz \cite{ref9}.

The atoms are trapped at the center of the glass cell with an outer size of 30~mm $\times$ 30~mm and a glass thickness of 3.5~mm. A microscope objective lens (Mitutoyo G Plan Apo 20X) is installed to take images of the atoms with high optical resolution. The objective lens has a numerical aperture of $NA$ = 0.28 and a long working distance of $WD$ = 30~mm. The lens also compensates for the glass thickness of 3.5~mm. These features of the lens allow an optical resolution of $\Delta_{\rm opt}$ = 1.2~$\mu$m for $\lambda$ = 671~nm to be easily realized by placing it outside the glass cell. A pixel size of $l_{\rm pix}$  = 13~$\mu$m and a measured magnification of $M$= 7.68 with an eyepiece lens of $f$ = 75~mm give the size of the visual field per pixel of $L_{\rm pix}=l_{\rm pix}/M=1.7$~$\mu$m. $L_{\rm pix} > \Delta_{\rm opt}$, so that $L_{\rm pix}$ limits the spatial resolution in this setup.

The probe laser beams are applied along the $y$ direction, and they are overlapped with the MOT laser beams with polarizations orthogonal to each other. A set of quarter-wave plates and a wire grid polarizer are inserted between the glass cell and the objective lens to transmit the probe beams and reflect the MOT laser beams \cite{ref10,ref11}. Although the objective lens does not compensate for the thickness of the quarter-wave plate and the wire grid polarizer, imaging of a test target confirmed that the spatial resolution is maintained.
While it is possible to evaluate actual spatial resolution by using cold atoms, for example, evaluation of atomic shot noises \cite{ref7.05}, we believe the spatial resolution evaluated by the test target.

The probe laser beams are focused in only the $x$ direction using a cylindrical lens. The 1/${\it e}^2$ beam diameters are ($d_x,d_z$) = (200~$\mu$m, 2~mm) at the position of the trapped atoms. The probe beams are slightly tilted in the $x$ direction, so that any reflected light on the surfaces of the optics is spatially separated from the optical path of the probe laser. Such reflections are finally blocked by a slit in front of the CCD camera. In this way, overlapping of the probe laser and their reflections is prevented and the emergence of interference fringes is suppressed.

\section{\label{crosssection}Calibration of the absorption cross-section}

\subsection{Principle}

The column density is given by dividing OD by the absorption cross-section $\sigma_{\rm abs}$.
In most cases, the effective value of $\sigma_{\rm abs}$ has a smaller value than the theoretical value $\sigma_{\rm abs}^0=3\lambda^2/2\pi$ \cite{ref5}.
Here, the calibration coefficient $\alpha$ is defined as $\sigma_{\rm abs}\equiv \sigma_{\rm abs}^0/\alpha$ with $\alpha \ge 1$.
We assume that $\alpha$ has a constant value which is independent of the probe intensity, pulse duration, atomic density, and other experimental parameters such as magnetic field.

We need to prepare some reference physical quantities to determine the value of $\alpha$.
One of the references that can be used is the local number density of an ideal non-interacting Fermi gas.
Experimentally, the local density of the trapped gas can be calculated from the column density $\bar{n}$ through the inverse Abel transform in an axially symmetric case: 
\begin{equation}
n(\rho,z)=-\frac{1}{\pi }\int_{\rho}^{\infty}\frac{1}{\sqrt{x^2-\rho ^2}}\frac{\partial \bar{n}(x,z)}{\partial x}dx.
\label{eq26}
\end{equation}
The relationship between the exact density $n$ calculated with $\sigma=\sigma_{\rm abs}^0/\alpha$ and $\tilde{n}$ calculated with $\sigma=\sigma_{\rm bas}^0$ is given by 
\begin{equation}
n(\rho,z)=\alpha \cdot \tilde{n}(\rho,z).
\label{nnbar}
\end{equation}
On the other hand, the local number density $n({\bf r})$ of an ideal Fermi gas trapped in a potential $U_{\rm trap}({\bf r})$ is theoretically determined under the local density approximation (LDA) as the EOS:
\begin{equation}
n(\mu({\bf r}), T)=-\frac{1}{\lambda_T^3 (T)} Li_{3/2} \left[ -{\rm exp}\left( \frac{\mu ({\bf r})}{k_B T} \right) \right],
\label{eqEOSdensity}
\end{equation}
with $\mu({\bf r})=\mu_0-U_{\rm trap}({\bf r})$, where $\lambda_T=\sqrt{2\pi \hbar^2/mk_B T}$ is the thermal wave length, $Li_s=\sum_{k=1}^{\infty}z^k/k^s$ is a polylogarithm function, and $k_B$ is the Boltzmann constant. 
From Eqs.~(\ref{nnbar}) and (\ref{eqEOSdensity}), we obtain
\begin{equation}
\tilde{n}(\rho, z)=-\frac{1}{\alpha}\frac{1}{\lambda_T^3 (T)} Li_{3/2} \left[ -{\rm exp}\left( \frac{\mu_0-U_{\rm trap}(\rho, z)}{k_B T} \right) \right].
\label{eqPrinciple1}
\end{equation}
Therefore, $\alpha$ can be determined along with $\mu_0$ and $T$ by fitting Eq.~(\ref{eqPrinciple1}) to the experimental data calculated from $OD$ with $\sigma_{\rm abs}^0$.
This method was used to calibrate $\sigma_{\rm abs}$ in Ref.~\cite{ref16}.
However, it is necessary to take the derivative of $\bar{n}(x,z)$ for the calculation of $n(\rho,z)$ from $\bar{n}(x,z)$ using Eq.~(\ref{eq26}).
This can sometimes significantly amplify any noise present in the experimental data.

For the sake of the SNR, comparison of the pressure rather than the number density was adopted in this work.
Theoretically, the pressure of an ideal trapped Fermi gas is given by
\begin{equation}
P(\mu({\bf r}), T)=-\frac{k_BT}{\lambda_T^3 (T)} Li_{5/2} \left[ -{\rm exp}\left( \frac{\mu ({\bf r})}{k_B T} \right) \right].
\label{eqEOSpressure}
\end{equation}
In previous work \cite{ref17, ref18}, the local pressure was simply related to the measured column density as
\begin{equation}
P(\rho=0,z)=\frac{m\omega_\rho^2}{2\pi}\int_{-\infty}^{+\infty}\bar{n}(x,z)dx,
\label{eq27}
\end{equation}
where $\omega_\rho$ is the radial trapping frequency.
$\alpha$ can then be determined using the same procedure as that when using the number density.
Since the pressure is given by integration of $\bar{n}(x,z)$, the local pressure can be obtained without amplification of the experimental noise.
However, Eq.~(\ref{eq27}) was derived under the assumption of a harmonic trapping potential.
Therefore, for better accuracy, we use below a general form of local pressure for an arbitrary axially symmetric potential $U_{\rm trap} (\rho,z)$ beyond a harmonic approximation \cite{ref19,ref19.5}.

\subsection{Derivation of local pressure}

A method which can calculate the local pressure from the column density without assuming a harmonic potential was shown by F\'elix Werner \cite{ref19}.
Local pressure is given by the Gibbs-Duhem equation, $P(\mu,T)=\int_{-\infty}^{\mu}n(\mu',T) d\mu'$.
The variable of the integration is converted to a space variable using the LDA.
Thus, the Gibbs-Duhem equation becomes ($z$ is fixed throughout)
\begin{equation}
P(\rho)=\int_\rho^\infty n(\rho')\frac{\partial U_{\rm trap}}{\partial \rho}(\rho')d\rho'.
\label{eq28}
\end{equation}
Substituting Eq.~(\ref{eq26}) into Eq.~(\ref{eq28}) gives the relationship between the pressure and the column density $\bar{n}$:
\begin{equation*}
P(\rho)=-\frac{1}{\pi}\int_\rho^\infty d\rho' \int_{\rho'}^\infty dx\frac{1}{\sqrt{x^2-\rho'^2}}\frac{\partial \bar{n}(x)}{\partial x}\frac{\partial U_{\rm trap}}{\partial \rho}(\rho').
\end{equation*}
We can change the order of integration and integrate by parts:
\begin{equation}
P(\rho)=\frac{1}{\pi}\int_\rho^\infty dx~\bar{n}(x)\frac{\partial}{\partial x}\int_\rho^x d\rho' \frac{1}{\sqrt{x^2-\rho'^2}}\frac{\partial U_{\rm trap}}{\partial \rho}(\rho').
\label{eq29}
\end{equation}
The Leibniz integral rule, $\frac{\partial}{\partial x}\int_\rho^{x-\epsilon}d\rho'~f(x,\rho')=f(x,\rho'=x-\epsilon)+\int_\rho^{x-\epsilon}d\rho'\frac{\partial}{\partial x}f(x,\rho')$ for small $\epsilon>0$, gives
\begin{eqnarray*}
\frac{\partial}{\partial x}\int_\rho^{x-\epsilon}d\rho'\frac{1}{\sqrt{x^2-\rho'^2}}\frac{\partial U_{\rm trap}}{\partial \rho}(\rho') \\
=\frac{1}{\sqrt{x^2-(x-\epsilon)^2}}\frac{\partial U_{\rm trap}}{\partial \rho}(x-\epsilon)-\int_\rho^{x-\epsilon}d\rho'\frac{x}{(x^2-\rho'^2)^{3/2}}\frac{\partial U_{\rm trap}}{\partial \rho}(\rho') \\
=\frac{1}{\sqrt{x^2-\rho^2}}\frac{\partial U_{\rm trap}}{\partial \rho}(x-\epsilon)+\int_\rho^{x-\epsilon}d\rho'\frac{\rho'\frac{\partial U_{\rm trap}}{\partial \rho}(x-\epsilon)-x\frac{\partial U_{\rm trap}}{\partial \rho}(\rho')}{(x^2-\rho'^2)^{3/2}}.
\end{eqnarray*}
By taking the limit $\epsilon \rightarrow+0$, it becomes
\begin{eqnarray}
\frac{\partial}{\partial x}\int_\rho^x d\rho' \frac{1}{\sqrt{x^2-\rho'^2}}\frac{\partial U_{\rm trap}}{\partial \rho}(\rho') \nonumber \\
=\frac{1}{\sqrt{x^2-\rho^2}}\frac{\partial U_{\rm trap}}{\partial \rho}(x)+\int_\rho^x d\rho'\frac{\rho'\frac{\partial U_{\rm trap}}{\partial \rho}(x)-x\frac{\partial U_{\rm trap}}{\partial \rho}(\rho')}{(x^2-\rho'^2)^{3/2}}.
\label{eq30}
\end{eqnarray}
From Eqs.~(\ref{eq29}) and (\ref{eq30}), the local pressure is obtained in the form \cite{ref19,ref19.5}
\begin{eqnarray}
P(\rho,z)=\nonumber\\
\frac{1}{\pi}\int_{\rho}^{\infty}dx~{\bar n}(x,z)\left[ \frac{\frac{\partial U_{\rm trap}}{\partial \rho}(x,z)}{(x^2-\rho^2)^{1/2}} + \int_{\rho}^x d\rho '~\frac{\rho '\frac{\partial U_{\rm trap}}{\partial \rho}(x,z)-x\frac{\partial U_{\rm trap}}{\partial \rho}(\rho ',z)}{(x^2-\rho '^2)^{3/2}} \right].\nonumber\\
\label{eq31}
\end{eqnarray}
The term in the square brackets in Eq.~(\ref{eq31}) can be calculated precisely because the trapping potential $U_{\rm trap}$ is analytically given by the beam waists, power of the ODT, and the magnetic curvature.
Therefore, the local pressure can be calculated by a simple integration of the column density with little influence from experimental noise as with the case of Eq.~(\ref{eq27}).

This formula can be readily modified for the case of elliptic symmetry, where every physical quantity is a function of $\rho=\sqrt{x^2+\eta^2(z)y^2}$ and $z$, with $\eta(z)$ being the ellipticity of the trapping potential.
In this case, the inverse Abel transform given in Eq.~(\ref{eq26}) is modified as \cite{ref21}
\begin{equation}
n(\rho,z)=-\frac{\eta(z)}{\pi}\int_{\rho}^{\infty}\frac{1}{\sqrt{x^2-\rho^2}}\frac{\partial \bar{n}(x,z)}{\partial x}dx.
\label{eq43}
\end{equation}
As a result, Eq.~(\ref{eq31}) can be modified to
\begin{eqnarray}
P(\rho,z)=\nonumber\\
\frac{\eta(z)}{\pi}\int_{\rho}^{\infty}dx~{\bar n}(x,z)\left[ \frac{\frac{\partial U_{\rm trap}}{\partial \rho}(x,z)}{(x^2-\rho^2)^{1/2}} + \int_{\rho}^x d\rho '~\frac{\rho '\frac{\partial U_{\rm trap}}{\partial \rho}(x,z)-x\frac{\partial U_{\rm trap}}{\partial \rho}(\rho ',z)}{(x^2-\rho '^2)^{3/2}} \right].\nonumber\\
\label{ellip}
\end{eqnarray}
The analytical forms of Eq.~(\ref{ellip}) for our trap configuration and the harmonic approximation are derived in \ref{sec:Analytical} and \ref{sec:HarmonicApproximation}, respectively.

\subsection{Examination of data analysis}

\begin{figure}[tb!]
 \centering
 \includegraphics{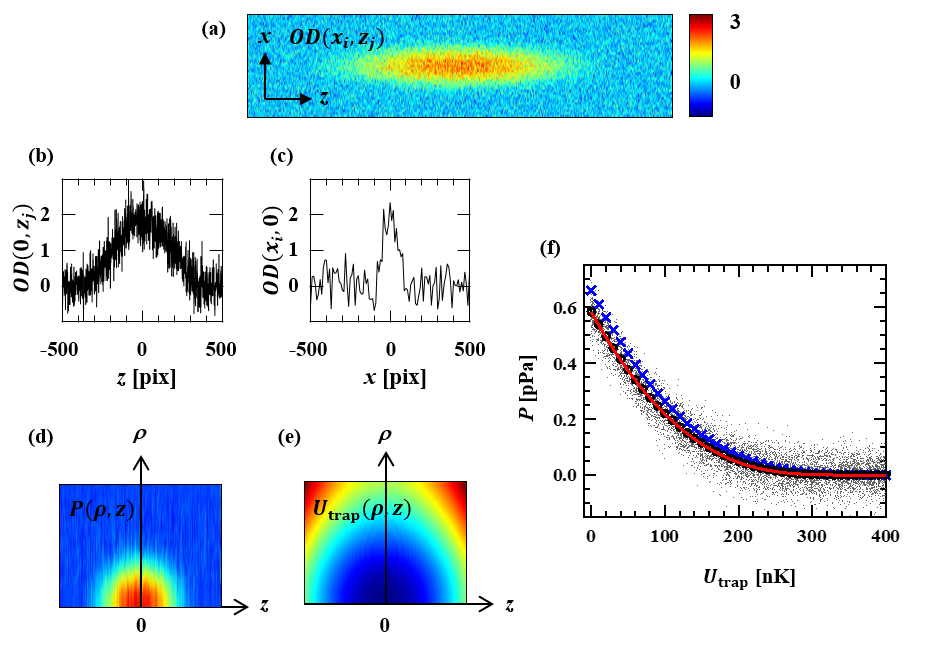}
 \caption{\label{fig5}
Examination of data analysis. (a) OD of an ideal Fermi gas trapped in the potential prepared numerically by using Eq.~(\ref{eq48}).
(b) Cross-section of OD at $x=0$, and (c) that at $z=0$.
(d) Local pressure $P(\rho,z)$ calculated using Eq.~(\ref{eq44}).
(e) The trapping potential $U_{\rm trap} (\rho,z)$.
(f) Scatter plots show $P$ as a function of $U_{\rm trap}$ obtained from (d) and (e) at each position, where the open circles show the averaged pressure.
The cross marks show the averaged pressure calculated under the harmonic approximation using Eq.~(\ref{eq45}).
The solid red curve shows the correct values given by Eq.~(\ref{eq49}) .
}
\end{figure}

The validity of the data analysis was examined. The column densities $\bar{n}(x,z)$ obtained by experiment have only 10 discrete data points from the peak to the tail in the $x$ direction. The experimental data also include noise. Similar data were numerically prepared and the data analysis based on Eq.~(\ref{eq44}) was examined to determine whether the correct pressure is obtained. The test column density was prepared using a thermodynamic function of the number density,
\begin{equation}
\bar{n}(x_i,z_j)=-\frac{1}{\lambda_T^3 (T)} \int_{-\infty}^{+\infty}dy~Li_{3/2} \left[ -{\rm exp}\left( \frac{\mu_0-U_{\rm trap} (x_i,y,z_j)}{k_B T} \right) \right],
\label{eq48}
\end{equation}
Random noises were also added to the column density over all the pixels of the CCD with same amplitude in order to roughly simulate the practical experimental data which has noises.

Test data of $\bar{n}(x_i,z_j)$ were prepared using Eq.~(\ref{eq48}) with the gas parameters $\mu_0/k_B=300$~$\mu$K and $T=20$~nK, the same potential parameters as the actual experiment of $P_{\rm ODT}=45.1$~mW, $w_{0x}=43.5$~$\mu$m, $w_{0y}=46.9$~$\mu$m, and $\omega_{\rm mag}=2\pi \times 6.96$~Hz, and the noise factor $SNR=5$, at 200$\times$1000 pixels with $L_{\rm pix}=1.7$~$\mu$m.
The trap depth is $U_{\rm{depth}}/k_B=850$~nK.
Since $\mu_0+k_BT \ll U_{\rm{depth}}$, the trap is deep enough to confine the degenerate Fermi gas.
Figure~\ref{fig5}(a) shows the prepared sample data, where $OD(x_i,z_j)=\sigma_{\rm abs}^0 \bar{n}(x_i,z_j)$, and Figs.~\ref{fig5}(b) and (c) show the cross-section along the axial direction $OD(x_i=0,z_j)$ and the radial direction $OD(x_i,z_j=0)$, respectively.
Fig.~\ref{fig5}(d) shows the local pressure $P(\rho,z)$ calculated using Eq.~(\ref{eq44}), and Fig.~\ref{fig5}(e) shows the trapping potential $U_{\rm trap} (\rho,z)$.

When $P(\rho \neq 0,z)$ is calculated in Eq.~(\ref{eq44}) by numerical integration with an equal interval of $\Delta t=L_{\rm pix}$, column density values are required at positions of $x=\sqrt{t_i^2+\rho^2}$, which do not exactly match the CCD addresses.
Therefore, in this analysis, these were calculated by interpolating data to the required positions.

Fig.~\ref{fig5}(f) shows a scatter plot of $P$ as a function of $U_{\rm trap}$ obtained at each position.
The averaged values calculated at a given height of $U_{\rm trap}$ are shown as open circles.
The same data were also analyzed under the harmonic approximation using Eq.~(\ref{eq45}), and the averaged values are shown as cross marks.
The data were compared to the correct values given by a thermodynamic function of pressure,
\begin{equation}
P(U_{\rm trap}; T,\mu_0)=-\frac{k_B T}{\lambda_T^3 (T)} Li_{5/2} \left[-{\rm exp}\left( \frac{\mu_0-U_{\rm trap}}{k_B T} \right) \right].
\label{eq49}
\end{equation}
The solid red curve shows the correct values. The pressures calculated using Eq.~(\ref{eq44}) are the correct values, while those calculated using Eq.~(\ref{eq45}) show deviations from these values. Therefore, the harmonic approximation causes systematic errors in calculations of the pressure.
The typical experimental data have an SNR better than 5, as shown in Fig.~\ref{fig4}(b).
Hence, it was concluded that this data analysis works correctly for the present experimental data.

\subsection{Determination of the absorption cross-section}

\begin{figure}[tb!]
 \centering
 \includegraphics{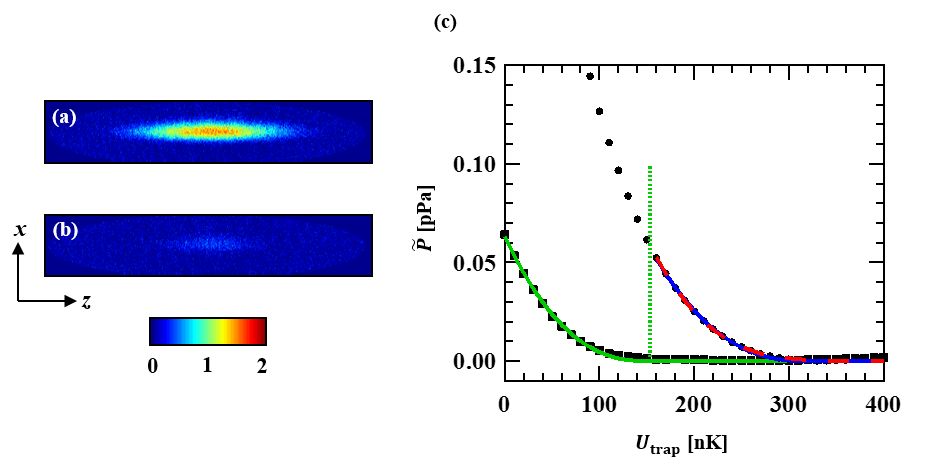}
 \caption{\label{fig6}
Calibration of the absorption cross-section $\sigma_{\rm abs}$. (a) OD of the majority, and (b) that of the minority in a highly imbalanced unitary Fermi gas. (c) Calibration of the absorption cross-section $\sigma_{\rm abs}$ using the nominal pressure $\tilde{P}$, which was calculated from Eq.~(\ref{eq44}) with $\bar{n}=OD/\sigma_{\rm abs}^0$, and the thermodynamic function for the ideal Fermi gas. The solid squares and solid circles indicate $\tilde{P}$ for the minority and the majority, respectively. The green solid curve is a fit to the minority using Eq.~(\ref{eq51}) to estimate the region of the minority. The vertical dotted green line shows the estimated edge of the minority. The dashed red curve and the solid blue curve show the fits to the majority outside of the minority using the EOS for an ideal Fermi gas at $T \neq 0$ (Eq.~(\ref{eq50})) and $T=0$ (Eq.~(\ref{eq51})), respectively.
}
\end{figure}

In this experiment, a region of degenerate ideal Fermi gas was prepared in a highly imbalanced unitary Fermi gas at 832.18~Gauss, as demonstrated in Ref.~\cite{ref16}. Figures~\ref{fig6}(a) and (b) show absorption images of state $|1\rangle$, which is the majority, and $|2\rangle$, which is the minority, in an imbalanced unitary Fermi gas. The nominal pressure $\tilde{P}$ is defined as the pressure given by Eq.~(\ref{eq44}) with $\bar{n}=OD/\sigma_{\rm abs}^0$.
The nominal pressure averaged over the region with a given potential height of $U_{\rm trap}$ for the majority and the minority is shown in Fig.~\ref{fig6}(c).
The value of $\tilde{P}$ and the exact value of $P$ follow the simple relationship $P(U_{\rm trap}; T,\mu_0)=\alpha \tilde{P}(U_{\rm trap})$. Therefore, $\alpha$ is determined along with $\mu_0$ and $T$ by fitting the nominal pressure to a function,
\begin{equation}
\tilde{P}(U_{\rm trap})=-\frac{1}{\alpha}\frac{k_B T}{\lambda_T^3 (T)} Li_{5/2} \left[-{\rm exp}\left(\frac{\mu_0-U_{\rm trap}}{k_B T}\right) \right].
\label{eq50}
\end{equation}
If $\mu_0/k_B T \gg 1$, then it can be approximated by the zero temperature limit:
\begin{equation}
\tilde{P}(U_{\rm trap})=\frac{1}{\alpha}\frac{1}{15\pi^2} \left(\frac{2m}{\hbar^2} \right)^{3/2} \left[ {\rm max}(\mu_0-U_{\rm trap},0) \right]^{5/2}.
\label{eq51}
\end{equation}

First, the region of the minority was estimated under the assumption of the zero-temperature limit.
We fit $\tilde{P}$ of the minority to Eq.~(\ref{eq51}) with two fitting parameters, $\alpha$ and $\mu_0$.
The determined $\mu_0$ was defined as $\mu_0^{\rm minority}$, which gives the region of the minority. The $\alpha$ determined by this fitting is not the correct calibration coefficient, because the minority is not an ideal Fermi gas but is interacting with the majority. In the region of $U_{\rm trap} > \mu_0^{\rm minority}$, the majority can be considered as single-component ideal fermions because they do not overlap with the minority.
Therefore, the calibration coefficient was determined by fitting $\tilde{P}$ of the majority at $U_{\rm trap} > \mu_0^{\rm minority}$ to Eq.~(\ref{eq50}) with three fitting parameters, $\alpha$, $\mu_0$, and $T$.
The fitting result is shown as the dashed red curve in Fig.~\ref{fig6}c.
The same experiment was repeated 50 times to decrease the statistical errors.
After averaging, the parameters $\alpha=2.0(2)$, $\mu_0/k_B=300(20)$~nK, and $T=20(10)$~nK were obtained.
Since the majority satisfies the condition of $\mu_0/k_B T \gg 1$, Eq.~(\ref{eq51}) is also available to evaluate $\alpha$.
In order to improve the fitting errors, we evaluated $\alpha$ by fitting Eq.~(\ref{eq51}) to the majority with two fitting parameters, $\alpha$ and $\mu_0$.
The fitting result is shown as the solid blue curve in Fig.~\ref{fig6}c.
The two fitting results by Eqs.~(\ref{eq50}) and (\ref{eq51}) overlap with each other, as shown in Fig.~\ref{fig6}(c).
As a result, the absorption cross-section was determined to be $\alpha= 2.06(9)$ within 4~\% uncertainty.

Our purpose is determination of a coefficient to convert from the OD to the column density.
While the value has a larger value than the theoretical values, it never obstruct the future experiments.
It can be caused by imperfection of the polarization of the probe laser, residual detuning caused by the Doppler shift, and the laser spectrum.
The influence of the fluorescence from the atoms is negligible for $NA=0.28$.

We note that there is a different method to determine the calibration coefficient $\alpha$ for homogeneous ideal fermions, which was developed in Ref.~\cite{ref21}.
The value $\alpha$ can be fixed by checking the relationship among compressibility, density, and pressure.

\section{\label{sec:Analytical}The analytical form of local pressure for a hybrid trap of an optical dipole trap and a magnetic trap}

Here we introduce an analytical form of Eq.~(\ref{ellip}) for the present trap configuration \cite{ref19.5}.
In the experiment, the trapping potential is given by a combination of an ODT ($U_{\rm ODT}$) and a magnetic trap ($U_{\rm mag}$) \cite{ref9}.
As noted in Sec.~\ref{Imaging}, the magnetic trap contributes only to the $z$ direction.
The trapping potential can then be well approximated by
\begin{equation}
U_{\rm trap} (x,y,z)=U_{\rm ODT} (x,y,z)+U_{\rm mag} (z).
\label{eq32}
\end{equation}
The magnetic trap is given by the simple form of a one-dimensional harmonic potential:
\begin{equation}
U_{\rm mag} (z)=\frac{m}{2}\omega_{\rm mag}^2 z^2.
\label{eq42}
\end{equation}
Note that the form of $U_{\rm mag}(z)$ is irrelevant in the pressure formula (\ref{ellip}).
On the other hand, the shape of the ODT is given by the spatial distribution of the laser intensity $I_{\rm ODT} (x,y,z)$ and a coefficient $C_{\rm ODT}$, as \cite{ref20}
\begin{equation}
U_{\rm ODT} (x,y,z)=-C_{\rm ODT} (I_{\rm ODT} (x,y,z)-I_{\rm ODT} (0,0,0)),
\label{eq33}
\end{equation}
where $I_{\rm ODT} (x,y,z)$ for an elliptic Gaussian beam is given by
\begin{equation}
I_{\rm ODT} (x,y,z)=\frac{2P_{\rm ODT}}{\pi w_x (z) w_y (z) } {\rm exp}\left[-2\left(\frac{x}{w_x (z)}\right)^2-2\left(\frac{y}{w_y (z)}\right)^2 \right],
\label{eq34}
\end{equation}
and the constant is
\begin{equation}
C_{\rm ODT}=\frac{3\pi c^2}{2\omega_0^3} \left(\frac{\Gamma}{\omega_0-\omega_{\rm ODT} }-\frac{\Gamma}{\omega_0+\omega_{\rm ODT} }\right).
\label{eq35}
\end{equation}
Beam waists in the $x$ and $y$ directions are
\begin{eqnarray}
w_x (z)=w_{0x} \sqrt{1+(z/z_x )^2 },\label{eq36} \\
w_y (z)=w_{0y} \sqrt{1+(z/z_y )^2 }. \label{eq37}
\end{eqnarray}
The Rayleigh lengths in the $x$ and $y$ directions are
\begin{eqnarray}
z_x=\frac{\pi w_{0x}^2}{\lambda_{\rm ODT}},\label{eq38} \\
z_y=\frac{\pi w_{0y}^2}{\lambda_{\rm ODT}}, \label{eq39}
\end{eqnarray}
and the effective Rayleigh length is defined as
\begin{equation}
z_R=\frac{\sqrt{2} z_x z_y}{\sqrt{z_x^2+z_y^2}}.
\label{eq40}
\end{equation}
Here, $\omega_0=2\pi c/\lambda_0$ and $\omega_{\rm ODT}=2\pi c/\lambda_{\rm ODT}$ are the frequencies of the atomic transition and that of the ODT laser, respectively.
The second term of Eq.~(\ref{eq33}) is added to make $U_{\rm ODT} (0,0,0)=0$.
In this case, the ellipticity of the trap is given by $\eta(z)=w_x (z)/w_y (z)$. 
Substituting Eq.~(\ref{eq32}) into Eq.~(\ref{ellip}) gives the local pressure in the following expression \cite{ref19.5}:
\begin{eqnarray}
P(\rho,z)=C_{\rm ODT} \frac{8P_{\rm ODT}}{\pi^2 w_x^2 (z) w_y^2 (z)} {\rm exp} \left[ -2\left( \frac{\rho}{w_x (z)} \right)^2 \right] \nonumber \\
\times \int_0^\infty dt~\bar{n}\left( x=\sqrt{t^2+\rho^2 },z \right)  \left( 1-2\sqrt{2} \frac{t}{w_x (z) } D\left( \sqrt{2} \frac{t}{w_x (z)} \right) \right),
\label{eq44}
\end{eqnarray}
where $D(x)$ is the Dawson function defined as $D(x)=e^{-x^2} \int_0^x e^{t^2} dt$.

\section{\label{sec:HarmonicApproximation}The analytical form of local pressure under the harmonic approximation}

In the case of $|x|\ll w_{0x}$, $|y| \ll w_{0y}$, and $|z| \ll z_R$, the harmonic approximation gives the approximated form of Eq.~(\ref{eq34}) around the origin:
\begin{equation}
I_{\rm ODT}^{\rm HO}(x,y,z)=\frac{2P_{\rm ODT}}{\pi w_{0x}w_{0y}}\left[ 1-2\left( \frac{x}{w_{0x}} \right)^2-2\left( \frac{y}{w_{0y}} \right)^2-\left( \frac{z}{z_{R}} \right)^2 \right].
\label{eq41}
\end{equation}
Under the harmonic approximation, pressure is approximated by
\begin{equation}
P^{\rm HO} (\rho,z)=C_{\rm ODT}\frac{8P_{\rm ODT}}{\pi^2 w_{0x}^2 w_{0y}^2 } \int_0^\infty dt \bar{n} \left(x=\sqrt{t^2+\rho^2},z\right).
\label{eq45}
\end{equation}
At $\rho=0$, it becomes
\begin{equation}
P^{\rm HO} (\rho=0,z)=C_{\rm ODT}\frac{4P_{\rm ODT}}{\pi^2 w_{0x}^2 w_{0y}^2}\int_{-\infty}^{+\infty} dx \bar{n}(x,z).
\label{eq46}
\end{equation}
The trapping frequencies are given by $\omega_x=\sqrt{\frac{4 U_0}{mw_{0x}^2}}$ and $\omega_y=\sqrt{\frac{4 U_0}{mw_{0y}^2}}$ with the depth of the potential, $U_0=C_{\rm ODT}\frac{2P}{\pi w_{0x} w_{0y}}$, so that Eq.~(\ref{eq46}) becomes
\begin{equation}
P^{\rm HO} (\rho=0,z)=\frac{m\omega_x \omega_y}{2\pi} \int_{-\infty}^{+\infty}dx \bar{n}(x,z),
\label{eq47}
\end{equation}
which is the same as Eq.~(\ref{eq27}).

\end{document}